\documentclass[aps,prl,reprint,superscriptaddress,amsmath]{revtex4-1}
\usepackage{graphicx}
\usepackage{textcomp}
\usepackage{upgreek}
\usepackage{hyperref}

\begin{document}


\title{Ultrafast Carrier Dynamics in VO$_2$ across the Pressure-Induced Insulator-to-Metal Transition}


\author{Johannes M. Braun}
\email[]{j.braun@hzdr.de}
\affiliation{Helmholtz-Zentrum Dresden-Rossendorf,  P.O. Box 510119, 01314 Dresden, Germany}
\affiliation{Technische Universit{\"a}t Dresden, 01062 Dresden, Germany}
\author{Harald Schneider}
\affiliation{Helmholtz-Zentrum Dresden-Rossendorf,  P.O. Box 510119, 01314 Dresden, Germany}
\author{Manfred Helm}
\affiliation{Helmholtz-Zentrum Dresden-Rossendorf,  P.O. Box 510119, 01314 Dresden, Germany}
\affiliation{Technische Universit{\"a}t Dresden, 01062 Dresden, Germany}
\author{Rafa\l{} Mirek}
\affiliation{Faculty of Physics, University of Warsaw, Pasteura 5, PL-02-093 Warsaw, Poland}
\author{Lynn~A. Boatner}
\affiliation{Oak Ridge National Laboratory, P.O. Box 2008, Oak Ridge, TN 37831, USA}
\author{Robert E. Marvel}
\affiliation{Vanderbilt University, Department of Physics and Astronomy, Nashville, TN 37235-1807, USA}
\author{Richard F. Haglund}
\affiliation{Vanderbilt University, Department of Physics and Astronomy, Nashville, TN 37235-1807, USA}
\author{Alexej Pashkin}
\email[]{a.pashkin@hzdr.de}
\affiliation{Helmholtz-Zentrum Dresden-Rossendorf,  P.O. Box 510119, 01314 Dresden, Germany}


\date{\today}

\begin{abstract}
We utilize near-infrared pump -- mid-infrared probe spectroscopy to investigate the ultrafast electronic response of pressurized VO$_2$. Distinct pump--probe signals and a pumping threshold behavior are observed even in the pressure-induced metallic state showing a noticeable amount of localized electronic states. Our results are consistent with a scenario of a bandwidth-controlled Mott-Hubbard transition. 
\end{abstract}


\maketitle


Vanadium dioxide (VO$_2$) exhibits a sharp insulator-to-metal transition (IMT) accompanied by a transformation from the monoclinic to rutile crystal structure above a critical temperature $T_\mathrm{c}=340\,$K \cite{Morin1959}. The strong coupling between electronic and lattice subsystems during the phase transition has attracted continuing interest for more than a half century \cite{Goodenough1971a,Imada1998,Basov2011}. The dimerized vanadium chains of the monoclinic phase suggests that a Peierls distortion underlies the insulating state, but electronic correlation leading to the carrier localization typical of a Mott insulator is also observed \cite{Biermann2005,Qazilbash2007,Weber2012,Huffman2017}. The IMT in VO$_2$ under the influence of temperature \cite{Koethe2006,Kim2006a,Qazilbash2007,Budai2014}, strain \cite{Gray2016a,Park2013,Aetukuri2013} and chemical substitution \cite{Marezio1972,Pouget1974} has been extensively studied. Some of these studies show that the electronic and structural phase transitions are separable within certain temperature ranges, pointing to a primary role for electron correlation as the driving mechanism for the IMT \cite{Kim2006a,Gray2016a}.

Application of external pressure offers an attractive way to distinguish the influence of the structural instability from the correlation effects on the IMT in VO$_2$. At sufficiently high pressures, VO$_2$ becomes metallic while keeping the dimerized monoclinic structure \cite{Arcangeletti2007,Marini2010,Marini2014,Baldini2016,Bai2015,Chen}. External pressure induces only an isostructural transformation of the lattice at room temperature \cite{Mitrano2012,Bai2015,Chen}. Thus, in this case the pressure-driven IMT should be dominated by changes in the electronic band structure which is necessarily different from that of the temperature-induced IMT. Therefore, it is highly desirable to unravel how the band structure of the monoclinic metallic phase VO$_2$ changes under high pressure. Unfortunately, in a high-pressure diamond anvil cell (DAC), conventional photoemission spectroscopy cannot be used, and X-ray absorption spectroscopy suffers from limited energy resolution \cite{Marini2014}.

Information about the electron and lattice dynamics in VO$_2$ can be obtained using time-resolved techniques that probe the evolution of the non-thermal IMT \cite{Cavalleri2001a,Cavalleri2004c,Cavalleri2005a,Hilton2007a,Kim2006a,Kuebler2007a,Pashkin2011,Cocker2012,Wall2012,Wegkamp2014a,Morrison2014,OCallahan2015,Huber2016}. In particular, time-resolved photoemission spectroscopy \cite{Wegkamp2014a} and ultrafast electron diffraction \cite{Morrison2014} have shown that a transient monoclinic but metallic phase can be induced via photoexcitation. A \textit{transient} monoclinic metallic phase has also been reported for pressurized VO$_2$ \cite{Hsieh2014a} using time-resolved reflectivity. 

Here we combine an ultrafast near-infrared pump and \mbox{mid-infrared} spectroscopy in a high-pressure DAC to investigate the non-equilibrium dynamics of the pressure-induced IMT in VO$_2$. Our results provide evidence that near-infrared pumping induces additional long-lived charge carriers -- even in the pressure-induced metallic phase. The utilized method of the non-degenerate nonlinear spectroscopy enables us to trace the evolution of localized, weakly localized and fully delocalized electronic states in VO$_2$ across the pressure-driven IMT, and to draw conclusions about the appropriate correlated band structure. 

We use VO$_2$ single crystals grown by thermal decomposition of V$_2$O$_5$ \cite{Budai2014,supplement}. The samples were polished to thicknesses of 20--30\,\textmu{}m and cut into pieces of around 100\,\textmu{}m in diameter. A single piece of VO$_2$ was mounted in an opposing-plate DAC. CsI powder was used as a pressure transmitting medium in order to ensure a direct contact between the sample surface and the front diamond anvil. The pressure inside the DAC was monitored via a standard ruby fluorescence method \cite{Mao1986}.

Our pump--probe setup is based on a Ti:sapphire laser amplifier system, providing 55\,fs long pulses centered at \mbox{$\lambda\approx800$\,nm (1.55\,eV)} with a repetition rate of 250\,kHz. A portion of the beam was used for the pumping branch, whereas the remaining part was utilized to generate probe pulses at \mbox{$\lambda=10$\,\textmu{}m} (0.12\,eV or 30\,THz) using difference frequency mixing between signal and idler pulses from a parametric amplifier. The pump and probe beams were focused non-collinearly on the sample inside the DAC down to a spot size of 30-50\,\textmu{}m (FWHM). We then measure the change in reflectivity $\Delta{}R$ of photo-excited VO$_2$ with respect to the reflectivity $R$ in the unexcited state. The small photon energy of the probe beam (well below the band gap energy $E_\mathrm{g}=0.6$\,eV of VO$_2$ at ambient conditions) ensures that the pump--probe signal is dominated by the response of free charge carriers \cite{Kuebler2007a,Pashkin2011}. 

\begin{figure}[t]
	\includegraphics{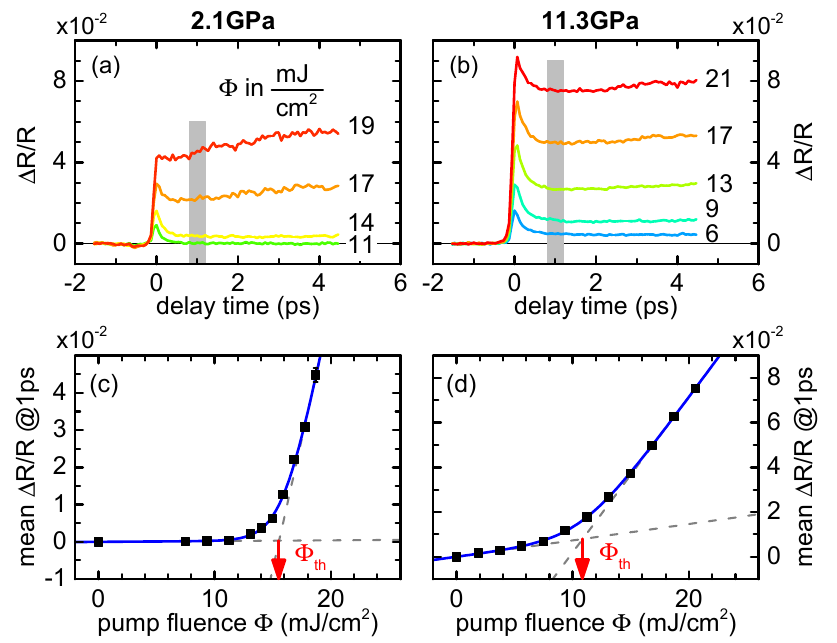}
	\caption{(color online). Calibrated pump--probe signals for different pump fluences $\Phi$ at \mbox{$T=300$\,K} and hydrostatic pressure of \mbox{(a) 2.1\,GPa} and \mbox{(b) 11.3\,GPa}. Time-averaged pump--probe signals for delay times between 0.8 and 1.2\,ps [highlighted as grey area in (a) and (b)] versus pump fluence for \mbox{(c) 2.1\,GPa} and \mbox{(d) 11.3\,GPa}. Black squares represent experimental data, the blue lines show corresponding fits. The crossing point of the asymptotes of the fits (dashed lines) determine the threshold fluence $\Phi_\mathrm{th}$.}
	\label{fig:1}
\end{figure}

Figure~\ref{fig:1}(a) shows the change of the pump--probe signal of VO$_2$ at 2.1\,GPa measured for different pump fluences. All of the curves show a quasi-instantaneous increase of the reflectivity -- limited only by the durations of the pump and the probe pulses. The onset is followed by a fast relaxation with a time constant of approximately 0.2\,ps \cite{supplement}. At low fluence (green trace), the pump--probe signal vanishes after about 1\,ps indicating a return to complete localization of the photo-excited charge carriers. At pump fluences above a threshold $\Phi_\mathrm{th}$ a persistant enhanced reflectivity reveals the creation of a metastable metallic phase [see the orange and red trace in Fig.~\ref{fig:1}(a)]. This state survives for hundreds of picoseconds in agreement with previous studies \cite{Becker1994,Hilton2007a,Brady2016}. All of these observations are fully consistent with the results for ambient conditions published previously \cite{Kuebler2007a}. The threshold fluence $\Phi_\mathrm{th}$ corresponds to the photoinjection of a critical density of free charge carriers that screen the Coulomb interaction and thus induce the collapse of the energy gap, thus leading to the metastable metallic state \cite{Kim2006a,Kuebler2007a,Wegkamp2014a}.

The volume of the metastable metallic phase and correspondingly the amplitude of the persistent pump--probe signal grows, when the incident pump fluence $\Phi$ is increased above the threshold $\Phi_\mathrm{th}$. This is related to the fact that the penetration depth of the mid-infrared probe and the sample thickness are much larger than the absorption length of the near-infrared pump; the photoexcitation switches only a relatively thin surface layer of the VO$_2$ crystal.

Remarkably, in contrast to the low-pressure regime, at elevated pressures the long-lived excited state is already observed at pump fluences below the threshold as shown in \mbox{Fig.~\ref{fig:1}(b)}. This becomes more obvious when we plot pump--probe signals averaged around 1\,ps (where the relaxation process is already completed) as a function of the pump fluence. Figures~\ref{fig:1}(c) and \ref{fig:1}(d) show such plots for the same pressures as for Fig.~\ref{fig:1}(a) and \ref{fig:1}(b), respectively. The threshold behavior is clearly seen in both cases. In contrast to previous works \cite{Kuebler2007a,Pashkin2011,Cocker2012}, the threshold is no longer well-defined by the crossing of the high-fluence asymptote with the x-axis, since finite signals are observed down to the lowest pump fluences. Therefore, we define $\Phi_\mathrm{th}$ as a crossing point (marked by red arrows) of the two asymptotes (dashed gray lines) illustrated in Fig.~\ref{fig:1}(c) and (d). Details of the bi-asymptotic fit function are given in \cite{supplement}.

\begin{figure}[t]
	\includegraphics{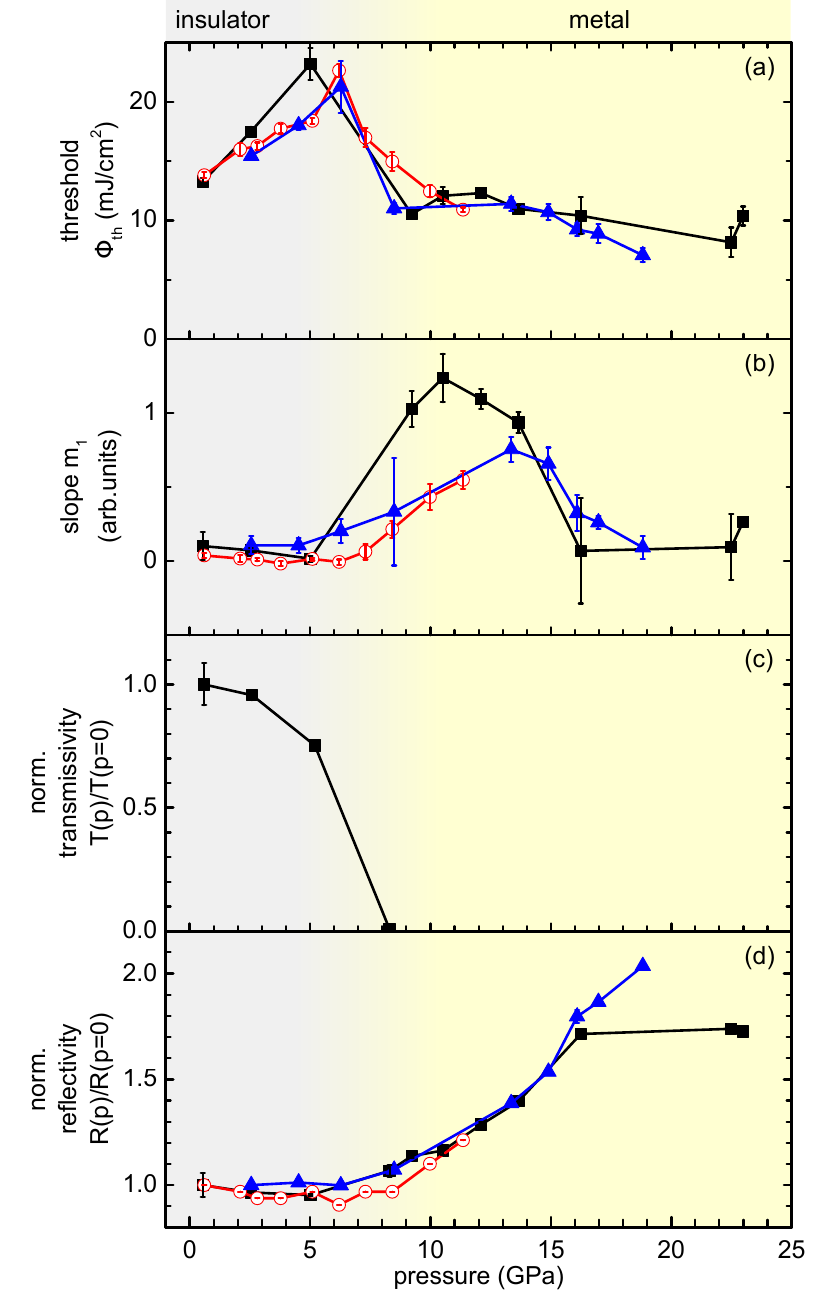}
	\caption{(color online). (a) Threshold fluence $\Phi_\mathrm{th}$ and \mbox{(b) slope $m_1$} obtained from the fitting as functions of pressure for three different $\mathrm{VO}_2$ crystals (corresponding to different colors and symbols). (c) Normalized linear transmissivity and (d) reflectivity of $\mathrm{VO}_2$ versus pressure.}
	\label{fig:2}
\end{figure}

The analysis of pump--probe traces at different pump fluences and various pressures reveals two main parameters that exhibit anomalous pressure behavior: (i) threshold fluence $\Phi_\mathrm{th}$ and (ii) slope of the low-fluence asymptote denoted as $m_1$ \cite{supplement}. Figure~\ref{fig:2} shows the pressure dependence of these key parameters together with the linear (unpumped) transmissivity and reflectivity at the probe photon energy.
Starting from ambient pressure, the threshold fluence $\Phi_\mathrm{th}$ monotonically increases up to a critical pressure $p_\mathrm{c}$ of 6-8\,GPa as shown in Fig.~\ref{fig:2}(a). A higher threshold fluence means that more free carriers have to be photoinjected to overcome the correlations and to induce the metastable metallic state. This trend contrasts sharply with the behavior of the thermally driven IMT where a noticeable decrease of the threshold fluence on approaching $T_c$ has been reported \cite{Pashkin2011,Cocker2012}. Most probably this difference is related to the stiffening of the dimerized lattice structure under pressure; that makes the vanadium dimers even more stable and thus raises the energy barrier to the metallic rutile phase. A stiffening of phonons observed in the pressure-dependent Raman measurements \cite{Arcangeletti2007,Bai2015} as well as enhanced dimerization of the vanadium sublattice seen in X-ray scattering \cite{Baldini2016} confirm this interpretation. Furthermore, the observed increase of $\Phi_\mathrm{th}$ with pressure is consistent with the recently reported growth of $T_c$ in VO$_2$ under pressure and the estimations of the corresponding increase in the latent heat of the IMT \cite{Chen}.

Around the critical pressure $p_\mathrm{c}$  we observe an anomalous drop of $\Phi_\mathrm{th}$. Remarkably, it coincides with the vanishing of the linear transmissivity [Fig.~\ref{fig:2}(c)], the start of the increase in the reflectivity [Fig.~\ref{fig:2}(d)] as well as the onset of a finite slope $m_1$ of the low-fluence asymptote [Fig.~\ref{fig:2}(b)]. This effect is highly reproducible and has been observed in independent measurements on three different VO$_2$ samples as presented in Fig.~\ref{fig:2}. We interpret the observed anomaly as a pressure-driven IMT in VO$_2$. Our results for the linear mid-infrared response agree with the work of Arcangeletti \textit{et al.} \cite{Arcangeletti2007} where this phenomenon was reported for the first time. The sharper drop in our transmission-versus-pressure data can be explained by the much larger thickness of our sample. The reflectivity starts to increase for $p>p_\mathrm{c}$ due to the presence of delocalized charge carriers and continues to grow as their density and correspondingly the plasma frequency become larger. In agreement with previous reports \cite{Arcangeletti2007,Marini2010,Mitrano2012,Bai2015}, our complementary high-pressure Raman measurements confirm that VO$_2$ samples also preserve the monoclinic crystal structure far above $p_\mathrm{c}$ \cite{supplement}.
 
Let us now discuss the nonlinear response of VO$_2$ in the pressure-induced metallic state. The similar shape of the pump-probe response beyond the critical pressure $p_\mathrm{c}$ [see Fig.~\ref{fig:1}(b)] indicates that the majority of vanadium $d$-electrons remain localized even in the pressure-driven metallic phase and a photoexcitation is still able to induce a phase with a higher conductivity. Moreover, the finite values of the threshold fluence [see Fig.~\ref{fig:2}(a)] suggest that the photo-induced metallization for $p>p_\mathrm{c}$ is governed by the same mechanism as for pressures below $p_\mathrm{c}$. The observed drop of $\Phi_\mathrm{th}$ above the IMT can be related to the finite pressure-induced density of free charge carriers that should lead to a partial screening of the Coulomb correlation. As a result, fewer photoinjected carriers are required to achieve the critical concentration that closes the band gap.

A further increase of pressure up to 23\,GPa causes gradual lowering of $\Phi_\mathrm{th}$ indicating that the density of free charge carriers grows with pressure. Nevertheless, it remains below the critical concentration necessary for complete suppression of the carrier localization. This is consistent with the behavior of the pressure-dependent reflectivity that monotonically increases up to the highest pressures -- suggesting that the plasma frequency of free charge carriers just slightly exceeds the frequency of our mid-infrared probe. Assuming an electron mass $m* \approx 2m_\mathrm{e}$ \cite{Qazilbash2007}, we estimate the density of free electrons to be roughly $2\times 10^{20}$\,cm$^{-3}$ -- still a factor five lower than the critical concentration of $10^{21}$\,cm$^{-3}$ needed for the photo-induced phase transition at ambient pressure \cite{Pashkin2011}. At high pressures, the critical concentration may be even higher. Thus, the equilibrium density of free carriers for $p>p_\mathrm{c}$ is still well below the critical limit, that could be the reason for the quite moderate decrease of the threshold beyond $p_\mathrm{c}$.

The pressure-induced IMT also leads to the onset of a non-vanishing slope $m_1$ of the low-fluence asymptote [see Fig.~\ref{fig:2}(b)]. This means that for $p > p_\mathrm{c}$ even pumping well below a threshold fluence can induce a metastable metallic phase with enhanced reflectivity, as illustrated in Fig.~\ref{fig:1}(b). In other words, a certain amount of long-lived free charge carriers directly proportional to the number of pump photons can be added without reaching the critical concentration for a band gap collapse. Thus, the appearance of a finite $m_1$ is expected to be a \emph{non-cooperative phenomenon} related to the photoexcitation of weakly localized states (WLS) located near the Fermi level.

\begin{figure}[t]
	\includegraphics{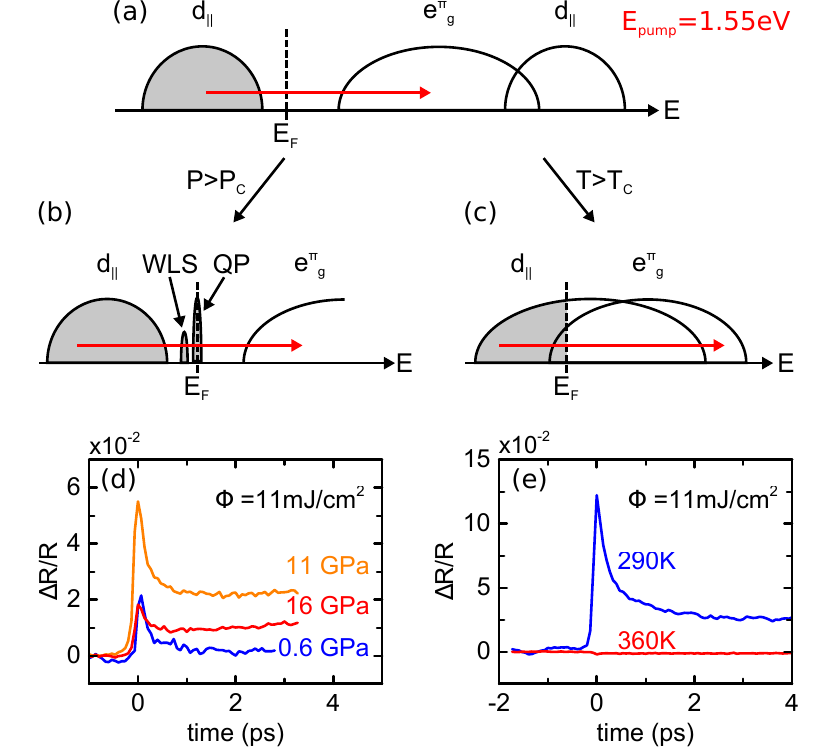}
	\caption{(color online). (a), (b) and (c) schematic energy bands of $\mathrm{VO}_2$ for (a) insulating monoclinic phase at ambient conditions ($T<T_\mathrm{c}$ and $p<p_\mathrm{c}$), (b) pressure-driven metallic phase and (c) temperature-driven metallic rutile phase. The qualitative drawing of the density of states (DOS) as function of the energy $E$ is restricted to the surrounding of the Fermi energy $E_\mathrm{F}$; in (b) appear weakly localized states (WLS) and a quasiparticle peak (QP), while the upper Hubbard band is above the shown energy scale. \mbox{(d) calibrated} pump--probe signal in reflectivity for different pressures $p$ at \mbox{$T=300$\,K} and pump fluence $\Phi=11\,\textrm{mJ}/\mathrm{cm^2}$. \mbox{(e) calibrated} pump--probe signal in reflectivity at ambient pressure and $\Phi=11\,\textrm{mJ}/\mathrm{cm^2}$ for $T<T_\mathrm{c}$ (blue line) and $T>T_\mathrm{c}$ (red line).}
	\label{fig:3}
\end{figure}

The observed behavior of the pump--probe response across the pressure-driven IMT can be understood on the basis of the tentative band diagrams depicted in Fig.~\ref{fig:3}(a)-(c). Figure~\ref{fig:3}(a) shows the band structure of VO$_2$ at ambient conditions established in previous studies \cite{Goodenough1971a,Biermann2005,Koethe2006}. The $t_{2g}$ vanadium orbitals overlapping along the $a_M$ axis \cite{Pashkin2011} form two relatively narrow bands usually denoted as $d_{\parallel}$. The low- and the high-energy bands correspond to bonding and antibonding combinations of electronic orbitals on a vanadium dimer, respectively. Due to the on-site Coulomb repulsion $U$ of $t_{2g}$ electrons, the lower and upper Hubbard bands are formed on the lower and upper energy ends of the $d_{\parallel}$ bands, respectively \cite{Biermann2005}. In our qualitative band diagrams, we present both bands as somewhat broader $d_{\parallel}$ bands. The band gap is formed between edges of the lower $d_{\parallel}$ band and the $e_g^\uppi$ band. Photoexcitation promotes the localized electrons from the $d_{\parallel}$ band into the high-energy delocalized states [red arrow in Fig.~\ref{fig:3}(a)] leading to the transient increase of reflectivity.

In our measurements, we observe photo-induced free charge carriers typical of the insulating VO$_2$ phase well above the pressure-induced IMT -- in stark contrast to the temperature-induced transition. This difference becomes clear by comparing the pump--probe signals measured on our VO$_2$ using the same excitation fluence for different pressures [Fig.~\ref{fig:3}(d)] and for different temperatures [Fig.~\ref{fig:3}(e)]. For temperatures above $T_c$, the metallic VO$_2$ phase possesses the rutile crystal structure without the dimerization. Thus, $d_{\parallel}$ bands become degenerate and cross the Fermi level together with other $t_{2g}$ bands [see Fig.~\ref{fig:3}(c)] resulting in a complete delocalization of all $3d$ electrons of vanadium ions. Therefore, the photoexcitation increases the electronic temperature, but it does not lead to an increase in the total number of free charge carriers. Correspondingly we observe a very weak negative change in $\Delta R/R$ for $T>T_\mathrm{c}$ as shown by the red trace in Fig.~\ref{fig:3}(e). In contrast to this, the pressure-induced metallic state of VO$_2$ is monoclinic \cite{supplement} suggesting that the $d_{\parallel}$ bands remain split. The drop of the threshold fluence leads to an enhanced photosusceptibility \cite{supplement}, and the sizable pump-induced $\Delta R/R$ for $p>p_\mathrm{c}$ signals shown in Fig.~\ref{fig:3}(d) indicate that a large part of the $3d$ electrons are localized, i.e., they occupy states in a band which does not cross the Fermi level. This assumption is supported by the X-ray absorption measurements by Marini \textit{et al.} \cite{Marini2014} which show that the spectral weight transfer needed to achieve the pressure-induced metallic monoclinic phase is much smaller than for the temperature-driven transition to the rutile phase. 

In addition, as discussed above, the appearance of finite $m_1$ values suggests the existence of weakly localized states WLS for $p > p_\mathrm{c}$. These conclusions result in the tentative band diagram shown in Fig.~\ref{fig:3}(b). It assumes that the band splitting is also preserved above the pressure-induced IMT, and the metallic conductivity originates from a spectral weight transfer to a narrow intragap quasiparticle peak (QP) at the Fermi level. Its low-energy satellite represents the WLS. At elevated pressures well above $p_\mathrm{c}$, the QP gains more spectral weight leading to higher metallic conductivity in line with the observed increase in reflectivity [Fig.~\ref{fig:2}(d)]. At the same time, the pump--probe signal $\Delta R/R$ becomes smaller since the photo-induced relative change in the density of free charge carriers decreases for a given pump fluence [see the decrease from orange to red trace in Fig.~\ref{fig:3}(d)]. Finally, $m_1$ vanishes at high pressures [Fig.~\ref{fig:2}(b)] indicating that the WLS merges with the QP.

We now discuss the scenario of the pressure-induced IMT in VO$_2$ via the suggested band diagram. Our data convincingly demonstrate that the dimerized monoclinic structure is preserved across the IMT and even becomes more stable under an initial pressure increase. Thus, the intimate coupling between the electronic and lattice subsystems characteristic of the temperature-driven IMT does not take place for the pressure-induced IMT which should have a predominantly electronic origin. The simplest scenario in this case is an IMT in the nondegenerate Hubbard model. Single-site dynamical mean field theory (DMFT) calculations for such model show that an increasing portion of spectral weight is transferred from the lower and upper Hubbard bands to a quasipartcle peak at the Fermi level when the effective correlation drops below a critical value, while the Hubbard bands persist \cite{Zhang1993,Khomskii2014}. The onset of the QP and the redistribution of spectral weight are governed by the ratio of the Coulomb repulsion $U$ and the hopping bandwidth $t$ \cite{Khomskii2014}. 

Application of pressure improves the overlap between the orbitals leading to increased bandwidth $t$ and reduced effective correlation $U/t$ that eventually results in a bandwidth-controlled Mott-Hubbard insulator-to-metal transition. Thus, this simple model is capable of explaining the existence of the localized states in the lower-energy band even above the IMT as suggested by our present study [Fig.~\ref{fig:3}(b)]. The only discrepancy is related to the existence of the WLS that are absent for the bandwidth-controlled IMT in the standard nondegenerate Hubbard model in single-site DMFT \cite{Zhang1993,Khomskii2014}.

We expect two possible reasons to be responsible for the appearance of WLS: (i) The WLS may be caused by lattice defects in the VO$_2$ crystal. Possibly their density increases notably due to the high strain imposed by external pressure \cite{OCallahan2015}. Initially delocalized electrons in the narrow QP are known to possess a high effective mass due to strong correlation effects \cite{Qazilbash2007,Khomskii2014} and, thus, may be localized and bound to lattice defects. With further increasing pressure and weaker correlation, the effective mass should strongly decrease such that the binding energy of the localized states may drop below the energy of thermal fluctuations at room temperature. As a result, the bound states will be ionized and the WLS peak will merge with the QP. (ii) The WLS may be an intrinsic feature of a realistic Hubbard model which goes beyond the single-site approximation and includes all relevant bands in VO$_2$ and must be solved using contemporary calculation techniques \cite{Biermann2005,He2016,Brito2016}. Unfortunately, to the best of our knowledge, no such modeling has yet been performed for a pressure-induced metallic phase of VO$_2$.       

In conclusion, we have observed a pressure-induced metallic monoclinic phase of VO$_2$ above a critical pressure $p_\mathrm{c}$ of 6-8\,GPa by using nonlinear pump--probe spectroscopy. The photo-induced response of VO$_2$ above the pressure-induced IMT is remarkably different as compared to the temperature-driven transition. This behavior agrees well with the scenario of a bandwidth-controlled Mott-Hubbard transition, where the strongly correlated metallic phase appears due to a spectral weight transfer from the Hubbard bands to delocalized states at the Fermi level. Thus, the application of external pressure provides a structural stability of VO$_2$ and reveales the purely electronic character of the insulator-to-metal phase transition.

\begin{acknowledgments}
	We thank A. Leitenstorfer and P.M. Oppeneer for fruitful discussions and support. This work was financially supported by the DFG (project 2113-1/1: A.P., J.M.B.).
	Research at the Oak Ridge National Laboratory for one author (L.A.B.) was supported by the U.S. Department of Energy, Office of Science, Basic Energy Sciences, Materials Sciences and Engineering Division.
	REM and RFH gratefully acknowledge funding from the National Science Foundation (DMR-1207507).
\end{acknowledgments}

\bibliography{bibfile}

\end{document}



\title{Supplemental material for "Ultrafast Carrier Dynamics in VO$_2$ across the Pressure-Induced Insulator-to-Metal Transition"}


\author{Johannes M. Braun}
\email[]{j.braun@hzdr.de}
\affiliation{Helmholtz-Zentrum Dresden-Rossendorf,  P.O. Box 510119, 01314 Dresden, Germany}
\affiliation{Technische Universit{\"a}t Dresden, 01062 Dresden, Germany}
\author{Harald Schneider}
\affiliation{Helmholtz-Zentrum Dresden-Rossendorf,  P.O. Box 510119, 01314 Dresden, Germany}
\author{Manfred Helm}
\affiliation{Helmholtz-Zentrum Dresden-Rossendorf,  P.O. Box 510119, 01314 Dresden, Germany}
\affiliation{Technische Universit{\"a}t Dresden, 01062 Dresden, Germany}
\author{Rafa\l{} Mirek}
\affiliation{Faculty of Physics, University of Warsaw, Pasteura 5, PL-02-093 Warsaw, Poland}
\author{Lynn~A. Boatner}
\affiliation{Oak Ridge National Laboratory, P.O. Box 2008, Oak Ridge, TN 37831, USA}
\author{Robert E. Marvel}
\affiliation{Vanderbilt University, Department of Physics and Astronomy, Nashville, TN 37235-1807, USA}
\author{Richard F. Haglund}
\affiliation{Vanderbilt University, Department of Physics and Astronomy, Nashville, TN 37235-1807, USA}
\author{Alexej Pashkin}
\email[]{a.pashkin@hzdr.de}
\affiliation{Helmholtz-Zentrum Dresden-Rossendorf,  P.O. Box 510119, 01314 Dresden, Germany}


\date{\today}

\begin{abstract}
\end{abstract}

\pacs{}

\maketitle


\section{Determination of the threshold parameters}
\label{sec:fitting}
In order to determine the threshold fluence we fit the fluence dependence of the pump--probe signal at the delay time of 1\,ps. For this purpose we use a phenomenological fit function with bi-asymptotic behavior:
\begin{equation}
\frac{\Delta{}R}{R}\left(\Phi\right)=c\ln\left[ae^{s_1\Phi}+\left(1-a\right)e^{s_2\Phi}\right],
\label{eq:1}
\end{equation}

with four independent positive fit parameters $c$, $a$, $s_1$ and $s_2$. In the limit of very large or very small fluences one of the exponential terms dominates, and the fit function demonstrates a nearly linear dependence on $\Phi$.

The threshold fluence $\Phi_\mathrm{th}$ is defined as a fluence at which the second derivative $\partial^2\frac{\Delta{}R}{R} / \partial^2 \Phi$ is maximal, i.e., the fit curve has the highest curvature. It can be shown that this condition implies the equality of both exponential terms in Eq.~(\ref{eq:1}) and, thus
\begin{equation}
\Phi_\mathrm{th}=\frac{1}{s_2-s_1}\ln\left(\frac{a}{1-a}\right).
\label{eq:2}
\end{equation}

In order to achieve a stable operation of the numerical fitting procedure an equivalent version of Eq.~(\ref{eq:1}) was used:
\begin{eqnarray}
\label{eq:3}
\frac{\Delta{}R}{R}\left(\Phi,\Delta{}t=1\,\mathrm{ps}\right)=\hspace{8em}\\c\ln\left[\frac{e^{s_3\Phi_\mathrm{th}}}{1+e^{s_3\Phi_\mathrm{th}}}e^{s_1\Phi}+\left(1-\frac{e^{s_3\Phi_\mathrm{th}}}{1+e^{s_3\Phi_\mathrm{th}}}\right)e^{\left(s_1+s_3\right)\Phi}\right],\nonumber
\end{eqnarray}

where the parameter $a$ that exponentially approaches 1 is replaced by the well-defined threshold fluence $\Phi_\mathrm{th}$. Also a new parameter $s_3=s_2-s_1$ is introduced for simplicity.

With the fit parameters $\Phi_\mathrm{th}$, $c$, $s_1$, $s_3$, the slope of the low-fluence asymptote $m_1$ can be derived by neglecting the second exponential term in Eq.~(\ref{eq:3}):

\begin{equation}
\label{eq:4}
m_1=cs_1
\end{equation}

The width of the transition between the low- and the high-fluence asymptotes can be estimated as the full width at half maximum (FWHM) of the second derivative of $\Delta{}R/R\left(\Phi\right)$ with respect to the fluence $\Phi$. One can show that

\begin{equation}
\label{eq:5}
\mathrm{FWHM}=\frac{2}{s_3}\ln\left(\frac{\sqrt{2}+1}{\sqrt{2}-1}\right)
\end{equation}

\begin{figure}
	\includegraphics{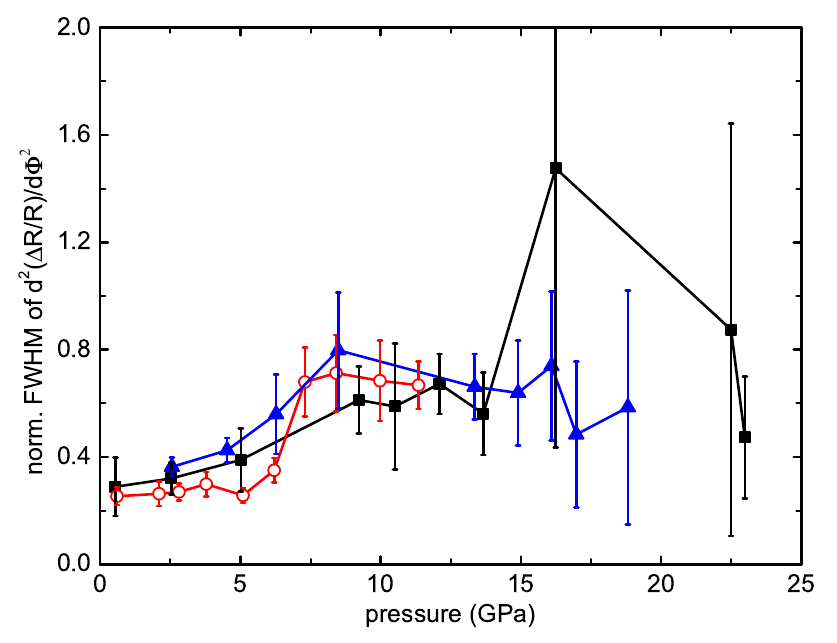}
	\caption{Pressure dependence of the FWHM parameter from Eq.~(\ref{eq:5}) normalized to the threshold fluence $\Phi_\mathrm{th}\left(p\right)$.\label{fig:SFig1}}
\end{figure}

Thus, the FWHM is determined just by the fit parameter $s_3$. Figure~\ref{fig:SFig1} shows how the unsharpness of the threshold behavior (ratio of the FWHM to the threshold fluence) changes with pressure. The pressure-induced metallization leads to a less sharp threshold behavior due to the onset of a finite slope $m_1$ and the drop in the threshold fluence. Nevertheless, the threshold behavior remains sufficiently sharp in the whole studied pressure range. The larger error bars at the high pressures are a consequence of weak pump--probe signals and low threshold values.

\section{Raman measurements on pressurized VO$_2$}
\begin{figure}
	\includegraphics{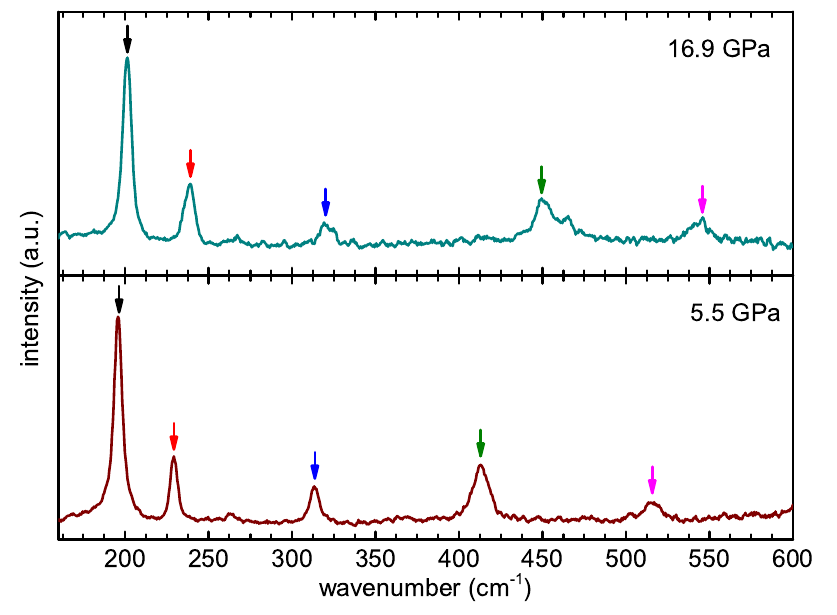}
	\caption{Raman spectra of VO$_2$ at 5.5\,GPa (brown) and 16.9\,GPa (cyan). Arrows mark the positions of the phonon modes plotted in Fig.~\ref{fig:SFig3}.\label{fig:SFig2}}
\end{figure}

Pressure-dependent Raman measurements were performed on a tiny, 10\,\textmu{}m thick single VO$_2$ crystal with 4:1 methanol-ethanol mixture as a pressure transmitting medium. Figure~\ref{fig:SFig2} shows Raman spectra at selected pressures below and above the IMT. The same set of phonon peaks can be observed up to the highest pressure indicating that the monoclinic structure is preserved. In contrast to the temperature-driven IMT, we do not observe a signature of the rutile metallic phase, where the sharp Raman peaks just disappear \cite{Schilbe2002}.

\begin{figure}
\includegraphics{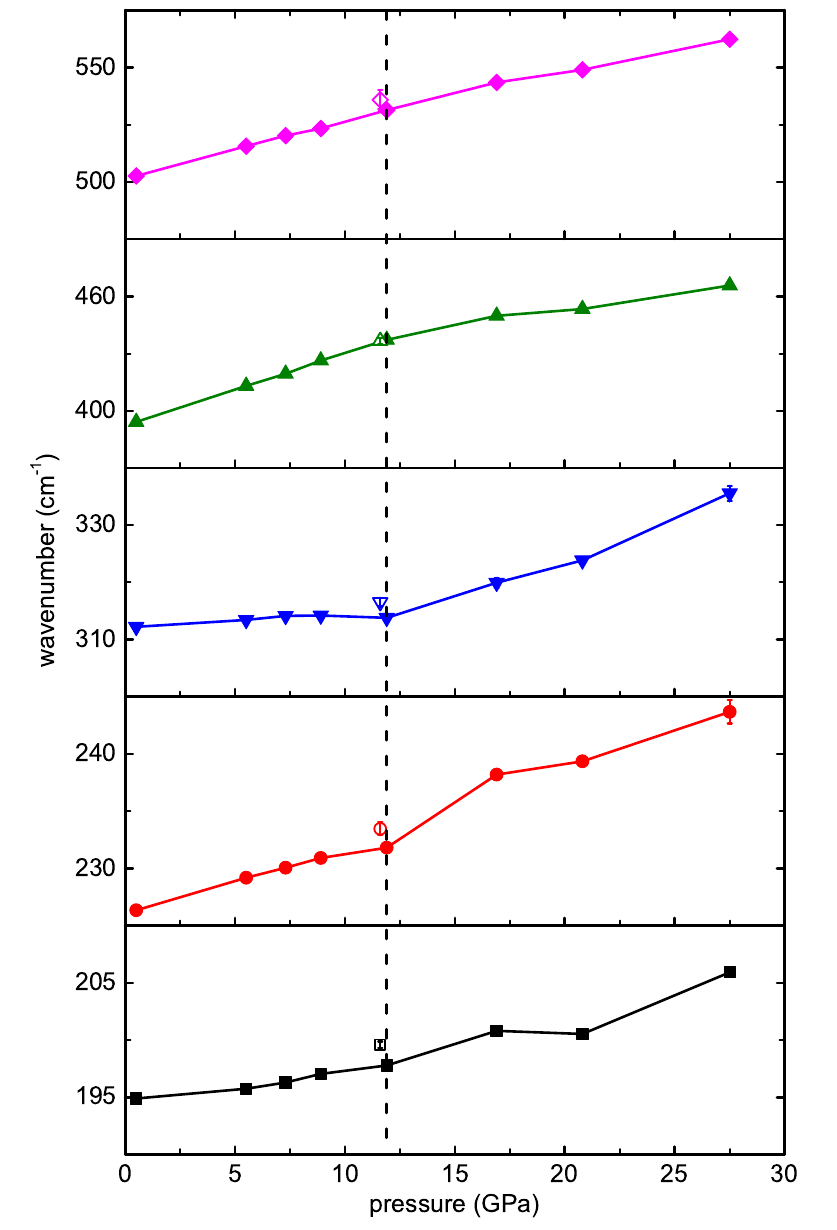}
\caption{Pressure dependence of selected Raman modes in VO$_2$. The open symbol corresponds to a Raman spectrum that was captured after the release of pressure. Color coding corresponds to the arrow markers in Fig.~\ref{fig:SFig2}. \label{fig:SFig3}}
\end{figure}

Figure~\ref{fig:SFig3} shows the frequencies of Raman modes as function of pressure. The kinks around 12\,GPa indicating an isostructural phase transition are in agreement with literature \cite{Arcangeletti2007,Marini2010,Mitrano2012,Bai2015}.

\section{VO$_2$ single crystals}
\begin{figure}
	\includegraphics[width=8.0cm]{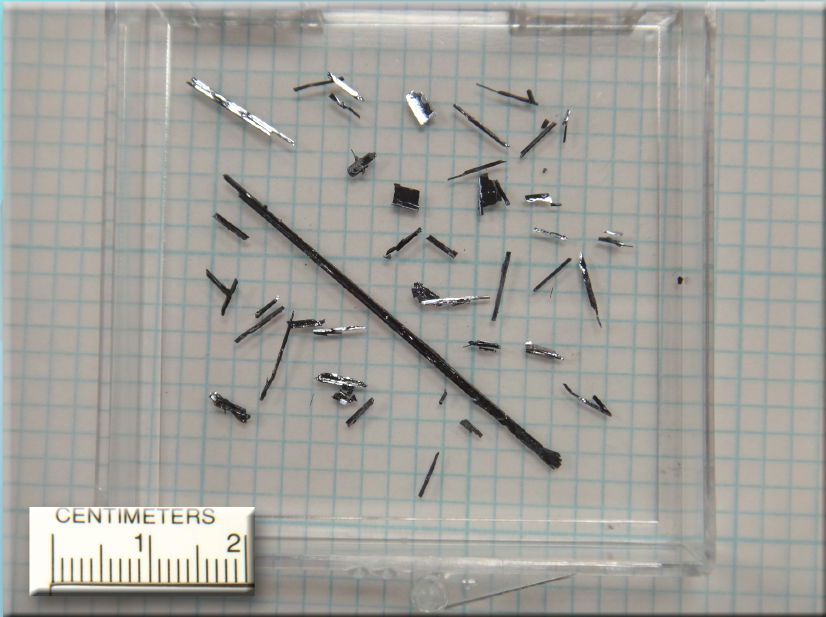}
	\caption{Single crystals of VO$_2$\label{fig:vo2crystals}}
\end{figure}

The single crystals of VO$_2$ [see Fig.~\ref{fig:vo2crystals}] were grown at the Oak Ridge National Laboratory using the methods described in Ref. \cite{Budai2014} and used in both Raman and pump--probe measurements.

\section{Negative pump--probe signal at small negative delay time}
Below the critical pressure $p_\mathrm{c}$ a small negative pump--probe signal is observed for negative delay times around 0.5\,ps getting more prominent for higher pump fluences [see Fig.~1(a)]. The origin of this feature is related to a pump-induced suppression of a portion of the probe pulse that is reflected from the backside of the sample. Initially the sample is transparent for the probe light, but when the portion reflected from the backside of the sample arrives the front surface excited by the pump pulse, its transmission is decreased by the layer of photo-excited free charge carriers. For fluences above the threshold the surface layer is metallic, and the suppression is very efficient. A rough estimation with a sample thickness of \mbox{20-30\,\textmu{}m} and refractive index of VO$_2$ around 3.2 \cite{Barker1966} results in the observed delay time of $-0.5\,$ps, which corresponds to temporal overlap of the probe pulse after the round trip through the sample with the incident pump pulse.

In the pressure-induced metallic phase the transmissivity of the sample vanishes and no light reflected from the backside of the sample can be detected. Therefore, the artifact at the negative delay time cannot be observed anymore for any pump fluence. Thus, the missing negative pump--probe signal beyond 6-8\,GPa is another evidence for a pressure-induced metallic phase [see Fig.~3(d)].

\section{Relaxation timescales in photo-excited VO$_2$}
\begin{figure}
	\includegraphics{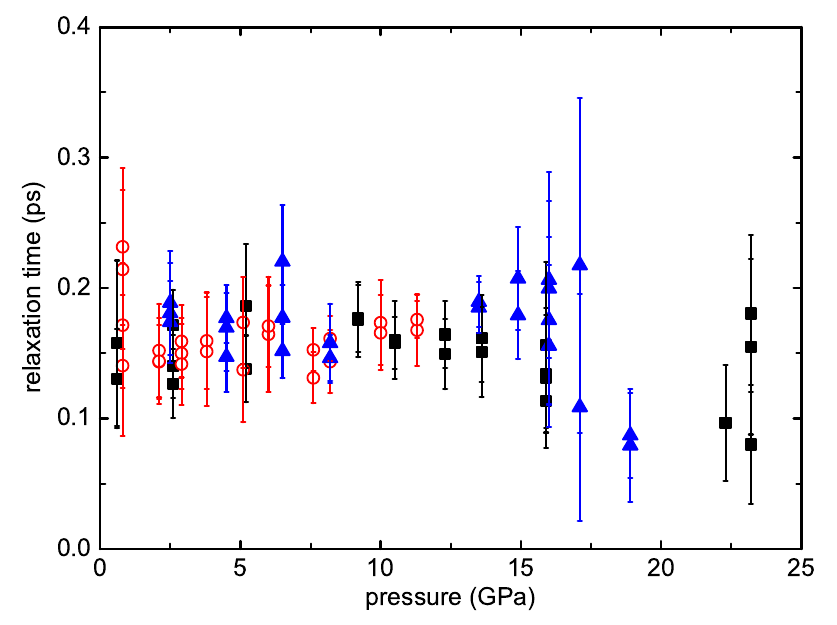}
	\caption{Fitted relaxation time constants for pump--probe traces corresponding to pump fluences within an interval of $5\,\textrm{mJ}/\mathrm{cm^2}$ below the threshold $\Phi_\mathrm{th}$ as function of pressure. \label{fig:SFig5}}
\end{figure}

The relaxation time constants of photo-excited charge carriers in pressurized VO$_2$ can be determined by mono-exponential fitting of the pump--probe traces at delay times between 0.05 and 1.0\,ps. In Fig.~\ref{fig:SFig5} the time constants obtained from pump--probe traces corresponding to pump fluences within an interval of $5\,\textrm{mJ}/\mathrm{cm^2}$ below the threshold $\Phi_\mathrm{th}$ are shown. The pump--probe traces measured at these fluences provide the most reliable estimation of the relaxation time: At lower fluences the signal-to-noise ratio is worse and at pump fluences beyond the threshold the long-lived signal affects the fitting. In general, we observe that the relaxation time is weakly dependent on the pump fluence.

The relaxation time constants for all three studied samples have pressure-independent values scattered in the range of 0.15 to 0.20\,ps. Previous study demonstrated similar values of the decay time \cite{Kuebler2007a}. There might by a weak pressure dependence with slightly shorter relaxation time constants at pressures beyond 16\,GPa. However, this could be an artifact caused by a decreased signal-to-noise ratio for the data taken at highest pressures where the signal amplitude becomes very low.

\section{Enhanced photosusceptibility in the pressure-induced metallic phase}
\begin{figure}
	\includegraphics{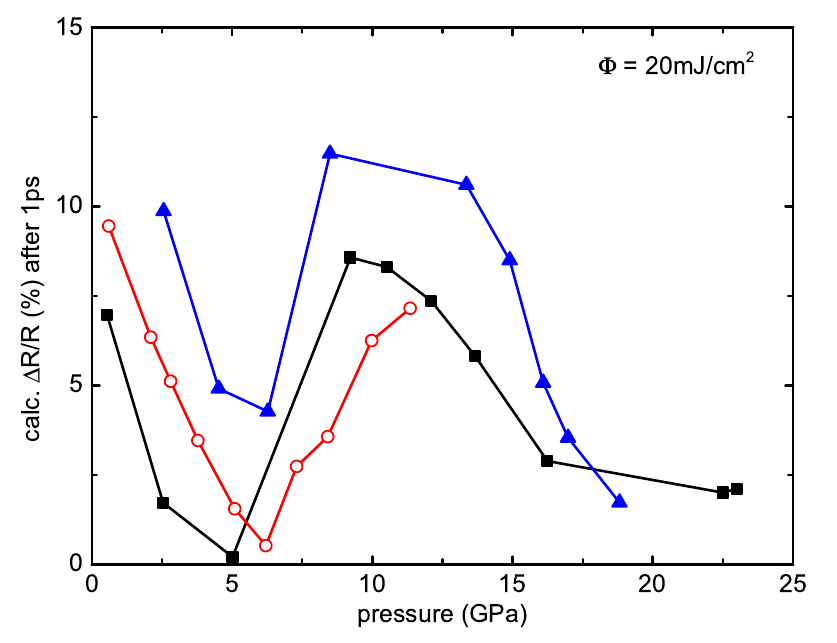}
	\caption{Pressure-dependence of the pump--probe amplitude for a constant pump-fluence $\Phi=20\,\textrm{mJ}/\textrm{cm}^2$. The traces are calculated with the parameters obtained from the threshold fitting and correspond to the three samples analyzed in Fig.~2. \label{fig:SFig6}}
\end{figure}

The amplitude of the pump--probe signal clearly increases at the critical pressure. This effect of enhanced photosusceptibility is illustrated in Fig.~\ref{fig:SFig6}, where the pump--probe amplitude for constant pump fluence \mbox{$\Phi=20\,\textrm{mJ}/\textrm{cm}^2$} is plotted as a function of pressure. Here we utilized the pressure-dependent parameters obtained from the threshold fitting to calculate the pump--probe signal at a delay time of 1\,ps for particular pressures.

Starting from ambient conditions, for all three samples the pump--probe signal decreases until the critical pressure is reached. This behavior as well as the sudden increase of the pump--probe amplitude beyond the critical pressure is a direct consequence of the first increasing and then dropping threshold fluence. The decrease of the pump--probe signal $\Delta R/R$ at even higher pressures can be explained by the rising number of pressure-induced free charge carriers, as then the photo-induced relative change in the density of free charge carriers lowers.

Enhanced photosusceptibility also was found for the temperature-driven IMT \cite{Hilton2007a}, but on the other side of the phase transition. While we observed this enhancement in the metallic phase, for the temperature-driven IMT it appears in the insulating phase. There the enhancement is due to the lowered thermodynamic barrier between the monoclinic and rutile phase when the monoclinic phase is heated \cite{Hilton2007a,Pashkin2011}.

\bibliography{bibfile}